\documentclass{aip-cp}
\usepackage[numbers]{natbib}
\usepackage[utf8]{inputenc}
\usepackage[T1]{fontenc}

\usepackage{graphicx}
\usepackage{aas_macros}
\usepackage{url}
\usepackage{subfigure}

\usepackage{color}

\begin{document}
%%%%%%%%%%%%%%%%%%% Title 
\title{Estimate Of The \textit{Fermi} Large Area Telescope Sensitivity To Gamma-ray Polarization}

\author[aff1]{Matteo Giomi\footnote{matteo.giomi@desy.de}~~~}%\corref{cor1}}
\author[aff1]{Rolf B\"{u}hler}
\author[aff2]{Carmelo Sgrò}
\author[aff3]{Francesco Longo}
\author[aff4]{W. B. Atwood}
\author[]{on behalf of the Fermi LAT Collaboration}

\affil[aff1]{Deutsches Elektronen-Synchrotron DESY, D-15738 Zeuthen, Germany}
\affil[aff2]{INFN-Pisa}
\affil[aff3]{INFN-Trieste, University of Trieste}
\affil[aff4]{University of California, Santa Cruz Institute for Particle Physics}
%\corresp[cor1]{matteo.giomi@desy.de}
\maketitle

%%%%%%%%%%%%%%%%%%% abstract
\begin{abstract}
Although not designed primarily as a polarimeter, the \textit{Fermi}-Large Area Telescope (LAT) has the potential to detect high degrees of linear polarization from some of the brightest gamma-ray sources. To achieve the needed accuracy in the reconstruction of the event geometry, low-energy ($\leq200$ MeV) events converting in the silicon detector layers of the LAT tracker have to be used. We present preliminary results of the ongoing effort within the LAT collaboration to measure gamma-ray polarization. We discuss the statistical and systematic uncertainties affecting such a measurement. We show that a $5\sigma$ minimum detectable polarization (MDP) of $\approx30-50\%$ could be within reach for the brightest gamma-ray sources as the Vela and Crab pulsars and the blazar 3C 454.3, after 10 years of observation. To estimate the systematic uncertainties, we stack bright AGN, and use this stack as a test source. LAT sensitivity to polarization is estimated comparing the data to a simulation of the expected unpolarized emission of the stack. We measure a 5$\sigma$ sensitivity limit corresponding to a polarization degree of $\approx37\%$. This is in agreement with a purely statistical estimate, suggesting that the systematic errors are likely to be small compared to the statistical ones.
%$5\sigma$ sensitivity is at the level of $\lesssim30\%$ polarization degree with 5 years of observation.
\end{abstract}

\section{Introduction}

%% define polarization, why it tells something about the source
Polarization is an intrinsic property of an electromagnetic wave. It describes the geometry of the oscillation of the electric field with respect to the wave's propagation direction. Since astronomical observations deals with a superposition of waves, a net polarization signal can emerge only in the presence of anisotropies in the emission (or absorption) region. Polarization carries important information on the astrophysical environment at the sources such as the geometry and orientation of the magnetic field, the location of the emitting region, and on the emission mechanism. The different science cases for polarization are reviewed in~\cite{revpol_pp, pol_and_pol_rev}.

%% origin of polarized gamma rays
Of the processes at the origin of high-energy gamma rays, only $\pi^{0}$ decay produces unpolarized radiation. Synchrotron emission is able to produce linearly polarized fluxes up to a high polarization degree of $\approx60\%-70\%$, given a uniform magnetic field at the source~\cite{radProc}. Compton scattering can both produce (in case of inverse Compton scattering) or reduce the degree of polarization of gamma-rays, with the  magnitude of the effect depending on the geometry of the interaction, see~\cite{ComptonPol} and references therein.

%% brief hystory of polarization in X (weisskopf, integral/IBIS), what could polarization in gamma tell us
X-ray polarization was first measured in the 1970s~\cite{weisskopf_a, weisskopf_b} for the Crab Nebula. More recently, polarization in soft ($\leq$ MeV) gamma rays has been measured for the Crab Nebula~\cite{crabpol_integral}, the the black hole binary system Cygnus X1~\cite{cygX1pol}, as well as some bright Gamma-ray burst, e.g.~\cite{integralGRB2013, grb_041219A_integral_a, grb_041219A_ibis_b}. Measuring gamma ray polarization at higher energies, in the pair-production regime, presents several challenges. Pair production is less sensitive than Compton scattering to polarization effect. Moreover, as the fluxes decrease steeply with energy, high-energy gamma-ray instrument are generally equipped with high-$Z$ converters to increase the effective area, at the expenses of a precise reconstruction of the event geometry.

%% gamma ray polarimetry future and why try with the LAT (pogolite, astromev, astrogam, harpo)
Although in the recent years there have been several proposal for future MeV missions (see, e.g.~\cite{astrogam, compair}) which also feature polarimetric capabilities, it is likely that no capable gamma-ray polarimeter will be operating in the upcoming years. For this reason, we investigate the capabilities of the \textit{Fermi} Large Area Telescope (LAT)~\cite{latpaper} to perform such a measurement. 

\section{Measurement of gamma-ray polarization with the LAT}

%% introduce the azimuthal modulation %In this regime, the orientation the plane of production of the $e^- / e^+$ pair is affected by the orientation of the linear polarization vector~\cite{polPP_olsen} of the photons. 
In the energy range of the LAT ($\gtrsim30$MeV) pair-production dominates the interaction of gamma-rays with matter. The cross section for this process changes depending on the relative orientation of the direction of the linear polarization vector of photon with respect to the plane where the $e^- / e^+$ pair is emitted~\cite{polPP_olsen}. As a result, a beam of linearly polarized photons produce a distribution of events that is modulated in the azimuthal angle $\psi$ of the plane of $e^- / e^+$ pair with respect to the direction of the lino of sight to the source:
\begin{equation}\label{eq:sig}
\frac{dN(\psi)}{d\psi}\propto1-A_{100}P\cos^2(\psi-\psi_{0})
\end{equation}
where $A_{100}$ is the amplitude of the modulation for 100\% polarized radiation, $P$ and $\psi_{0}$ are the polarization degree and angle of the incoming photons\footnote{The polarization degree is defined as the ratio between the intensity of polarized light over the total intensity of the incoming radiation. Polarization angles are conventionally measured from the celestial north going east.}. In Eq.\ref{eq:sig}, the observable quantity is the product $A_{100}P$. To reconstruct the degree of polarization, $A_{100}$ has to be calculated integrating the pair-production cross section in the kinematic phase space covered by the measurement. For this study we will assume $A_{100}\approx0.2$, as resulting from the integration of the cross section over the polar angles and recoil momenta, for a photon energy of 100 MeV~\cite{depaola}.

%% characteristic of the LAT tracker
The characteristics of the LAT tracker are detailed in~\cite{lat_tracker}. Each of the 16 modules (towers) of the the tracker is composed of a series of trays, each one carrying one converter foil of tungsten, and two single-sided silicon strip detectors (SSDs). Each SSD is 0.4\% radiation length in thickness and has a 228 $\mu$m strip pitch. The trays are stacked so that each tungsten foil is immediately followed by two SSDs with a 2 mm gap between them. The distance between the lower of these SSDs and the next tungsten foil is 30 mm. The top 12 trays of the towers have 'thin' converter foils of 2.7\% radiation length and constitute the 'FRONT' of the tracker, while the last 4 (the 'BACK' of the tracker) have thicker converter foils of 18\% radiation length.

\begin{figure}[h!]
\begin{minipage}[t]{0.5\linewidth}
    \includegraphics[width=0.83\linewidth]{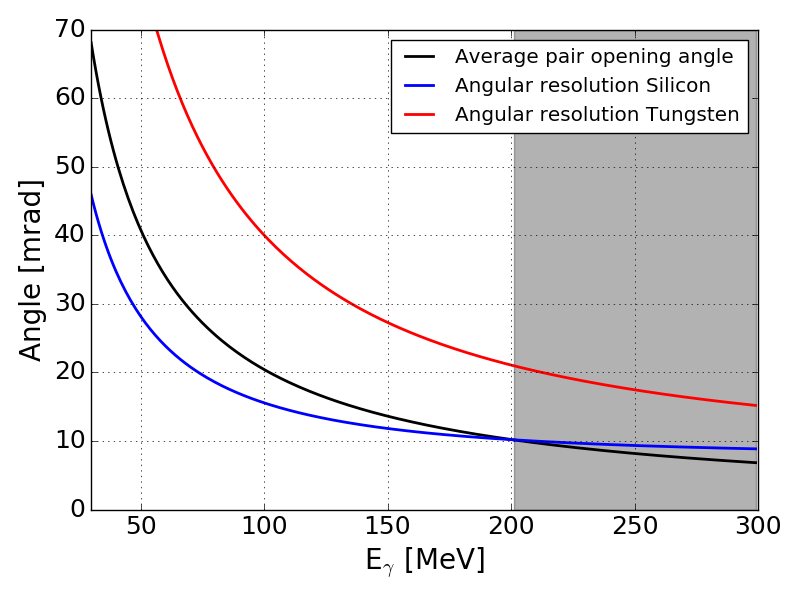}
%\textbf{A)} Comparison of the average pair opening angle (black line) and the achievable angular resolution for photons of different energies and converting either in the middle of a silicon layer (blue line), or in the middle of a tungsten foil (red line). The shaded region represent the energy range where the angular resolution is not sufficient to resolve the $e^- / e^+$ pair.\label{fig:polwhere}}
\centerline{\textbf{Fig. 1.a}}
\end{minipage}%
\hfill
\begin{minipage}[t]{0.5\linewidth}
    \includegraphics[width=0.83\linewidth]{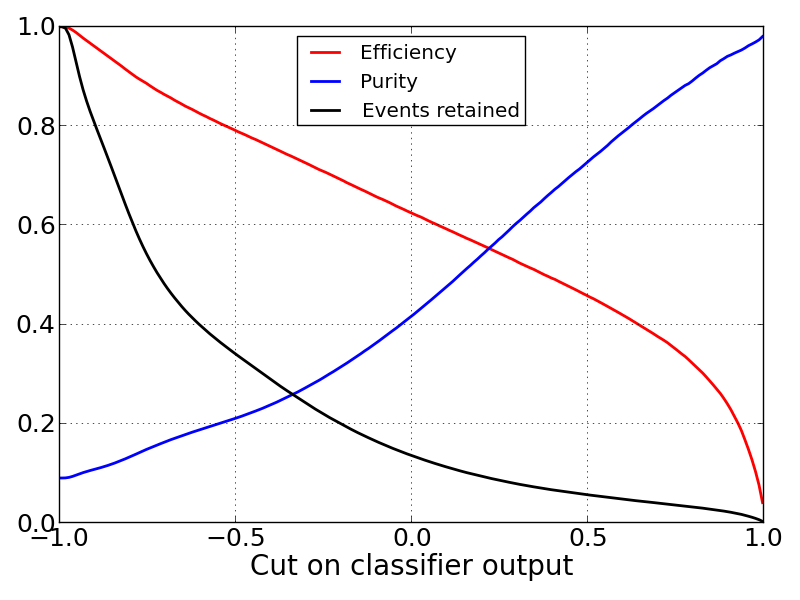}
%\textbf{B)} Performance of the selection of silicon converted events as a function of the cut on the classifier output variable: Efficiency (red), purity (blue), and fraction of retained events (black) as a function of the cut on classifier output.\label{fig:classifier}
\centerline{\textbf{Fig. 1.b}}
\label{fig:classifier}
\end{minipage}
\caption{
\textbf{a}) Comparison of the average pair opening angle (black) and the achievable angular resolution for photons of different energies and converting either in the middle of a SSD (blue), or in the middle of a tungsten foil (red). The shaded region represent the energy range where the angular resolution is not sufficient to resolve the $e^- / e^+$ pair. \textbf{b}) Performance of the classifier used to select silicon-converted events: efficiency (red), purity (blue), and fraction of retained events (black) as a function of the cut on classifier output $c_{BDT}$.\label{fig:polwhere_classifier}}
\end{figure}

%% Physic constraint on the measure
The angular resolution of the tracker is limited by the interplay of three angles: $\theta_{MS}$, the Multiple Coulomb Scattering (MS) angle~\cite{pdg}, $\theta_{op}$, the pair opening angle $\theta_{op}$, and $\theta_{geo}=7.6$ mrad, the geometrical limit to the LAT angular resolution due to the finite spacing between tracker strips. As shown in Fig.~\ref{fig:polwhere_classifier}.a, for photons converting in the tungsten converter foils of the tracker, the average opening angle $<\theta_{op}>\approx E_{\gamma}/m_{e}c^2$ is always smaller than the angular resolution, defined as $\theta_{res}=\sqrt{\theta_{MS}^2 + \theta_{geo}^{2}}$. For events that converts in the silicon conversions and for energies $\leq$200 MeV, the condition $\theta_{res}<\theta_{op}$ is satisfied and the azimuthal asymmetry could be measured.
  
%\begin{figure}[h]
%  \centerline{\includegraphics[width=0.465\linewidth]{polangles.png}}
%  \caption{Comparison of the average pair opening angle (black line) and the achievable angular resolution for photons of different energies and converting either in the middle of a silicon layer (blue line), or in the middle of a tungsten foil (red line). The shaded region represent the energy range where the angular resolution is not sufficient to resolve the $e^- / e^+$ pair. \label{fig:polwhere}}
%\end{figure}

To select silicon converted events, we use Boosted Decision Trees with gradient boosting (BDTG) from the \texttt{TMVA}\footnote{http://tmva.sourceforge.net/} package. The trees have been trained on a sample of Monte Carlo (MC) generated events of the P8V3\_SOURCE event class, the standard photon selection for point-like sources, in the energy range [30, 250] MeV. The sample includes $2\times10^4$ signal (silicon-converted) and $2\times10^5$ background (tungsten-converted) events. The classifier use 33 variables of the low-level \textit{Merit} event files. The response of the classifier evaluated on a separate sample of events shows no overtraining. In Fig.~\ref{fig:polwhere_classifier}.b the efficiency and purity of the classifier are shown as a function of the cut on the classifier output $c_{BDT}$. Using these response curves, we find the value of the cut on $c_{BDT}$ that maximize $N_{sig}/\sqrt{N_{bg}}$. This happens for $c_{BDT}>0.495$, corresponding to $\epsilon_{eff}=0.45$ and $\epsilon_{pur}=0.72$ and $f_{BDT}=5.7\%$ events remaining. 

%\begin{figure}[h]
%\centerline{\includegraphics[width=0.45\linewidth]{eff_pur.png}}
%\caption{Efficiency (red), purity (blue), and fraction of retained events (black) as a function of the cut on classifier output.\label{fig:classifier}}
%\end{figure}

\subsubsection{Minimum detectable polarization}

With less than $\approx6\%$ of the events passing the selection for silicon conversions, statistical uncertainties are expected to play a major role in limiting the LAT polarization sensitivity. The statistical sensitivity is estimated via the minimum detectable polarization, $\mbox{MDP}(p)$~\cite{mdp}, representing the minimum polarization degree that correspond to a modulation amplitude with less than a probability $p$ of being exceeded by chance in the absence of a signal:
\begin{equation}\label{eq:mdp}
\mbox{MDP}(p)=\frac{2}{A_{100}}\frac{\sqrt{-\ln(p)}}{R_{S}}\sqrt{\frac{R_{S}+R_{B}}{0.2T}}
\end{equation}
where $R_{S}$, $R_{B}$ are the rate of signal and background events, $T$ is the observation time that is scaled by a factor $0.2$ to account for the finite size of the LAT field of view. The rate of signal and background events after the analysis cuts are given by:
\begin{equation}\label{eq:sig_bg_rate}
R_{S}=0.95F^{src}A_{eff}f_{BDT}f_{op}\epsilon_{pur}
\qquad
R_{B}=[(1-\epsilon_{pur})0.95F^{src}+F^{diff}]A_{eff}f_{BDT}f_{op}
\end{equation}
were $\epsilon_{pur}$ and $f_{BDT}$ characterize the selection of silicon-converted events, and $f_{op}=0.7$ is the expected fraction of the events with large pair opening angle~\cite{OpenAngle}. $F^{src}$ is the source flux in the $[50, 200]$ MeV energy range, and $F^{diff}$ is the total flux of the diffuse emission within one $R_{PSF95}$ from the source in the same energy range. Throughout this study we will use Pass8~\cite{pass8} data, P8V3\_SOURCE event class including both FRONT and BACK events, and Instrument Response functions (IRFs) version P8V3\_SOURCE. The LAT effective area is averaged in the $[50, 200]$ MeV energy range and over incident angles $\theta \in [0^\circ, 60^\circ]$ weighting by a power-law flux of index -2, obtaining $A_{eff}\approx 1.9\times10^3$ cm$^2$. The 95\% containment radius of the Point Spread Function (PSF) is measured from the simulation of a point source resulting in $R_{PSF95}^{*}\approx11^\circ$ for all events, and $R_{PSF95}\approx7.6^\circ$ for silicon-converted events. 

\begin{figure}[h]
  \centerline{\includegraphics[width=0.5\linewidth]{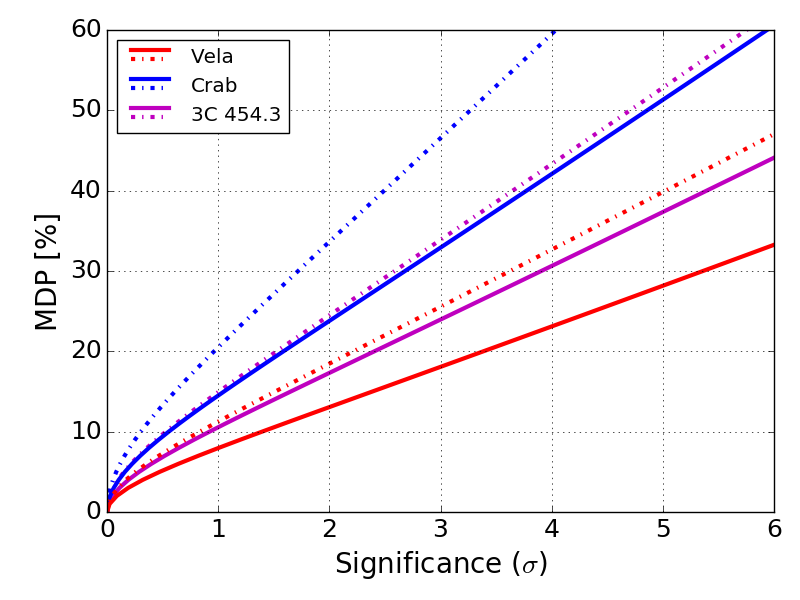}}
  \caption{Minimum detectable polarization at different significance levels for the Crab and Vela pulsar and 3C 454.3 assuming 10 years (solid lines) and 5 years (dash-dotted lines) of observation.\label{fig:statsig}}
\end{figure}

The MDP at different significance levels is presented in Fig.\ref{fig:statsig} for the three most promising sources: the Crab and Vela pulsars, and the blazar 3C 454.3. To produce these curves, we took $F^{source}$ from the \textit{Fermi} LAT Third Source Catalog (3FGL, ~\cite{3fgl}) and estimate $F^{diff}$ with the diffuse model (gll\_iem\_v06.fit\footnote{http://fermi.gsfc.nasa.gov/ssc/data/access/lat/BackgroundModels.html}). A $5\sigma$ MDP of $\approx30\%$ for 10 years of observation of the Vela pulsar. Expected values for polarization degrees of the synchrotron emission of pulsars varies between $\approx10\%-30\%$, with higher polarization degree for the off-pulse emission~\cite{pulsarPol}. Interesting is also the  case of the blazar 3C 454.3. Depending on the emission model blazar polarization can reach maximum values of $\approx70\%$ in case of hadronic and $\lesssim40\%$ for leptonic emission respectively, see e.g.~\cite{blazar_pol, blazarPol2}.

\section{Estimate of the systematics}

%% idea of the stack
To assess the level of the systematic errors we evaluate the precision with which a null signal can be measured. As no single gamma-ray source can be considered unpolarized with certainty, our test source will be made of a stack of Active Galactic Nuclei (AGN). Even if individually polarized, the polarization angles of the different sources are expected to be uncorrelated. Stacking many sources will wash out the residual polarization.

%% requirements of the stack.
We set up a toy model in order to decide the stack composition. Using the 3FGL and the diffuse emission templates, and observation parameters similar to those presented in the previous chapter, we evaluate the flux of each source and of the corresponding galactic diffuse background in the [50, 200] MeV energy range.
%% construction of the stack
We assume that all the point sources are highly polarized (polarization degree $P=60\%$) in a random direction. The Galactic diffuse emission is considered unpolarized. Each source will contribute to the total azimuthal distribution of the stack with a signal of the form of Eq.~\ref{eq:sig}, normalized to the number of counts expected from that source. To minimize BG contamination of the sample, AGN are added to the stack in order of $F^{src}/F^{diff}$~\footnote{Although 3C 454.3 is a good candidate for a real measure, it is still included in the stack. Even if very bright, it is not dominating the sample. Moreover, the 5-year duration of this analysis prevent any eventual polarization signal from 3C 454.3 to have any noticeable effect on the emission of the stack sources.}. To avoid overlap, sources within $20^\circ$ from any other stack member are discarded. For every new source added to the stack, the polarization signals from all the sources are summed, each one with a random phase $\psi_0$ and the the amplitude of the azimuthal modulation of the stack ($A_{stack}$) is measured. We repeat this sum $10^3$ times, each time randomizing the phases. The distribution of $A_{stack}$ is used to characterize the residual polarization of the stack. The process of growing the stack ends when the following requirements are fulfilled:
\begin{itemize}
\item Total flux from the stack sources $F^{stack}$ greater than $10^{-5}~\mbox{cm}^{-2}\mbox{s}^{-1}$
\item Low residual polarization: $<A_{stack}>$ is at least $2\sigma$ away from 2\%
\end{itemize}
%% the stack
The resulting stack consist of 25 sources. The stack total flux is $F^{stack}\approx1.9\times10^{-5}~\mbox{cm}^{-2}\mbox{s}^{-1}$ between 50 and 200 MeV and the diffuse flux in the same energy range, and within 6$^\circ$ from the sources is $\approx10^{-5}~\mbox{cm}^{-2}\mbox{s}^{-1}$.
%The total flux of these sources is $\approx1.3\times10^{-5}$ cm$^{-2}$s$^{-1}$, while the total diffuse emission within one $R_{PSF95}$ is of the order of $1.3\times10^{-5}$ cm$^{-2}$s$^{-1}$. With these values, the estimated statistical sensitivity for 5 years observation and a $5\sigma$ threshold is $P\approx28\%$, similar to the one we get for the Vela pulsar.

\subsubsection{Data and Monte Carlo simulation}

\begin{figure}[h]
  \centering
  \includegraphics[width=0.44\textwidth, height=5.5cm]{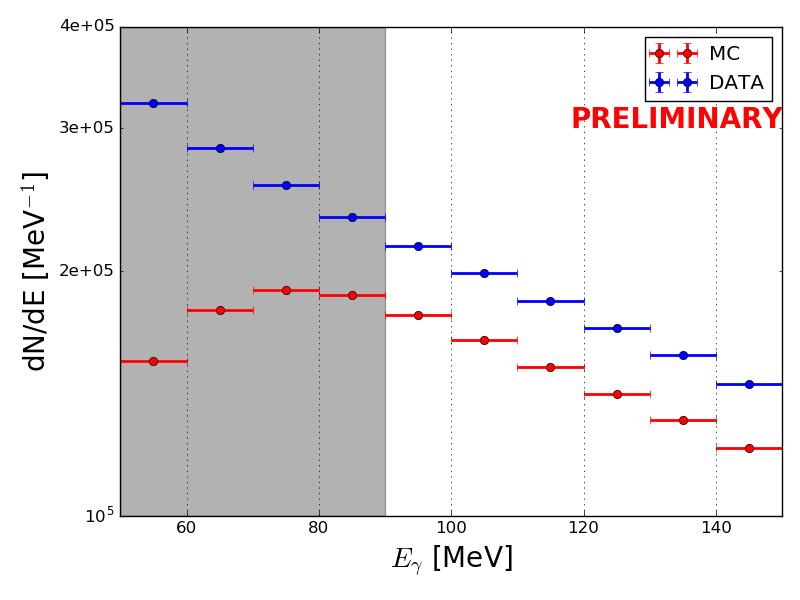}
  \hfill%
  %\hspace{1cm}
  \includegraphics[width=0.55\textwidth, height=5.5cm]{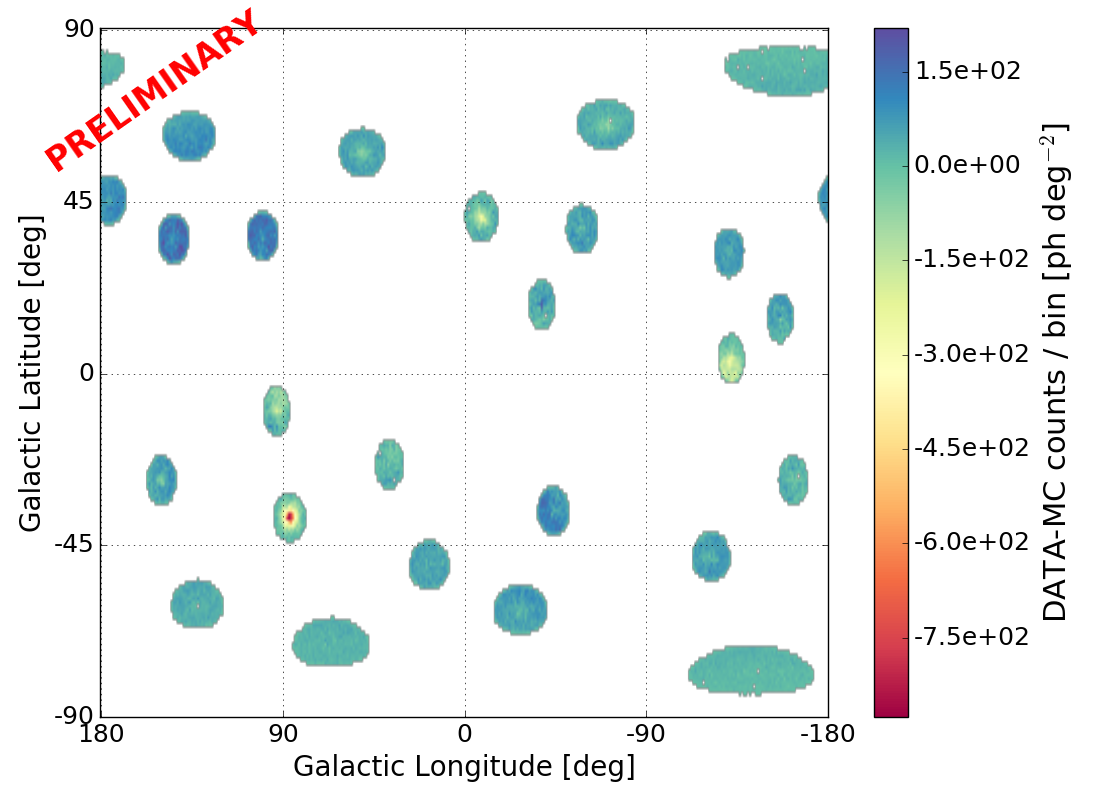}
  \caption{Data-MC comparison for data and simulated events with energies in [50, 150] MeV. Left panel: energy distribution of the events. Right: sky map of the Data-MC counts in a 6$^\circ$ regions around each source. The large residuals visible for 3C 454.3 (at l=86.12$^\circ$, b=-38.19$^\circ$) are due to the extreme variability of this source that lead to a over prediction of the quiescent flux. \label{fig:datamc}}
\end{figure}

This analysis covers roughly 5 years of observations, from 2008-08-04 to 2013-05-10 (MET 239560179 to 389884599). The simulation use the GLAST LAT Event Analysis Machine (\textit{Gleam}~\cite{gleam}) version 20-09-09. For the simulation, we use a realistic pointing history for the time range of interest, the models of the sky regions using 3FGL data for the point sources, and the recommended galactic diffuse and isotropic templates for the P8V3\_SOURCE event class. The simulation includes also all 3FGL point sources within $15^\circ$ to any stack source, if their flux is $>5\%$ of the AGN flux. The comparison of the energy and spatial distribution of simulated data and MC events within 6$^\circ$-radius circle around the sources is presented in Fig.~\ref{fig:datamc} for event energies in $[50, 150]$ MeV, and with zenith angle $<100^\circ$.

% caveats of this analysys
As shown in Fig.~\ref{fig:datamc}, the simulation under-predict the number of counts. The difference Data-MC increase as larger region around the sources are considered. The Data-MC difference below 90 MeV (shaded region in Fig.~\ref{fig:datamc} left) is primarily due to energy dispersion, as the the simulation of the Galactic diffuse emission starts at 60 MeV. In addition, an overall difference in the normalization of the MC counts with respect to the data can be seen above $\approx90$ MeV. The main reason for this is that we have not yet included the Earth Limb emission and faint point sources in the simulations. We expect none this to affect significantly the azimuthal distribution of the events\footnote{We tested how much the non-simulated Earth Limb emission affected the results applying different cuts on the Zenith angles of the events. The result shows no significant difference between a hard cut of 80$^\circ$ and a loose cut of 100$^\circ$.}. In the following we will consider events within 6$^\circ$ from the AGN of the stack and in the [90, 150] MeV energy range, where the data-MC agreement can be considered satisfactory up to the mentioned scaling factor, at the expenses of reduced statistics.

\subsubsection{Results and conclusions}

%% basic analysis
The azimuthal angle $\psi$ of each event is reconstructed using only the best (highest energy) track of the event. The second direction we will use to determine $\psi$ is determined by the source position. We therefore assume that all the recorded events comes from the point source, an assumption that is justified only in case of high signal to noise. With the vector $\hat{s}$ identifying the source position in the celestial sphere we define the unit vector $\hat{n'}$ as the projection of the north vector on the plane tangent to the sphere at the tip of $\hat{s}$. Be $\hat{t}$ the vector describing the best track of the event in the celestial sphere, and $\hat{t'}$ the projection of this vector on the tangent plane. The angle $\psi$ is then defined as:
\begin{equation}\label{eq:psi}
\psi=\arccos\left(\frac{\hat{t'}\cdot\hat{n'}}{|\hat{t'}|}\right)
\end{equation}

\begin{figure}[h]
  \centerline{\includegraphics[width=0.5\linewidth]{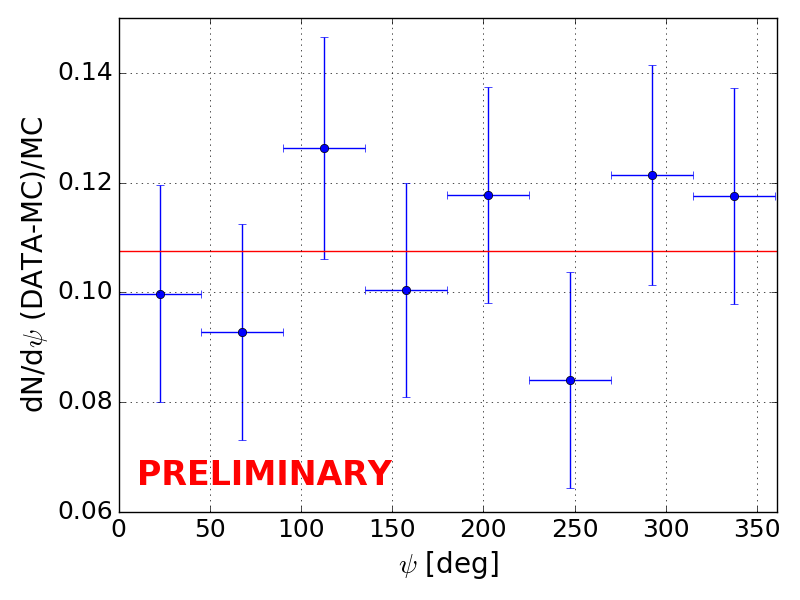}}
  \caption{Residual azimuthal distribution (DATA-MC)/MC for silicon-converted events in the energy range [90, 150] MeV. The red line is a constant term at $10.7\%$.\label{fig:azdist}}
\end{figure}

We compute the azimuthal distribution for silicon-converted events for data and for MC. The residual distribution, defined as (Data-MC)/MC is presented in Fig.~\ref{fig:azdist}. As visible, it is consistent with a constant (with a p-value of 76\%) within the statistical errors. This distribution can be used to estimate the global sensitivity of our analysis, including statistical (the stack is roughly as bright as Vela) and systematic uncertainties for silicon-converted events. Fitting this distribution with a signal of the form of Eq.~\ref{eq:sig} of increasingly larger amplitude, we find $A_{sens}$, the minimum amplitude of a signal that is not compatible with the residual azimuthal distribution at 5$\sigma$ confidence level. The sensitivity limit of this analysis is then $P_{sens}=A_{sens}/0.2$. With the event selection described above, we find $A_{sens}=7.3\%$ corresponding to $P_{sens}\approx37\%$. This result is in agreement with the purely statistical MDP for the stack of $\approx40\%$, suggesting that systematic uncertainties are likely to be small, and that the sensitivity is dominated by statistical errors.

%\subsubsection{Outlook and Conclusions}	    evaluate the LAT sensitivity for polarization measurement and provides
The analysis presented here demonstrates a method to evaluate the LAT sensitivity for polarization measurement and provides a first rough estimate of this sensitivity. It is therefore far from optimal. In particular a more complete sky model for the simulation, fitted to real data, should improve the data-MC agreement and allow to assess the systematics down to 50 MeV. A much larger MC production is also needed if one wants to eliminate the contribution of statistical errors on the MC, and improve the precision on the measure of the systematics. Dedicated \textit{Gleam} simulation that includes the polarization effect would also allow to verify and refine the event selection. An effort in this sense is ongoing in the \textit{Fermi}-LAT collaboration.

% aknowldegments
The \textit{Fermi} LAT Collaboration acknowledges generous ongoing support from a number of agencies and institutes that have supported both the development and the operation of the LAT as well as scientific data analysis. These include the National Aeronautics and Space Administration and the Department of Energy in the United States, the Commissariat \`a l'Energie Atomique 
and the Centre National de la Recherche Scientifique / Institut National de Physique Nucl\'eaire et de Physique des Particules in France, the Agenzia Spaziale Italiana and the Istituto Nazionale di Fisica Nucleare in Italy, the Ministry of Education, Culture, Sports, Science and Technology (MEXT), High Energy Accelerator Research Organization (KEK) and Japan Aerospace Exploration Agency (JAXA) in Japan, and the K.~A.~Wallenberg Foundation, the Swedish Research Council and the Swedish National Space Board in Sweden. Additional support for science analysis during the operations phase is gratefully acknowledged from the Istituto Nazionale di Astrofisica in Italy and the Centre National d'\'Etudes Spatiales in France.

%The \textit{Fermi}-LAT Collaboration acknowledges support for LAT development, operation and data analysis from NASA and DOE (United States), CEA/Irfu and IN2P3/CNRS (France), ASI and INFN (Italy), MEXT, KEK, and JAXA (Japan), and the K.A.~Wallenberg Foundation, the Swedish Research Council and the National Space Board (Sweden). Science analysis support in the operations phase from INAF (Italy) and CNES (France) is also gratefully acknowledged.

%%%%%%%%%%%%%%%%%%% References
\nocite{*}
\bibliographystyle{aipnum-cp}
\bibliography{polproc}

\end{document}